\documentclass[mathleft
]{an}
\usepackage{graphicx}
\usepackage{times}
\newcommand{\dg}{^{\circ}}
\begin{document}

\Pagespan{789}{}
\Yearpublication{2006}%
\Yearsubmission{2005}%
\Month{11}%
\Volume{999}%
\Issue{88}%

\title{Laboratory performances of the solar multichannel resonant scattering
       spectrometer prototype of the GOLF-New Generation instrument}

\author{S. Turck-Chi\`eze,  P.H. Carton, S. Mathur, J.-C. Barri\`ere, P. Daniel-Thomas, C. Lahonde-Hamdoun, \\R. Granelli, D. Loiseau, F. Nunio, Y. Piret\thanks{Corresponding author:
  \email{cturck@cea.fr, pcarton@cea.fr}\newline}
\and  J.M. Robillot\thanks{from the Bordeaux Observatory in France, presently retired}}
\titlerunning{Performances of GOLFNG prototype}
\authorrunning{S. Turck-Chi\`eze et al.}
\institute{DSM/IRFU/SAp, CEA/Saclay, 91191 Gif-sur-Yvette Cedex, France  and Laboratoire AIM, CEA/DSM -- CNRS - Universit\'e Paris Diderot -- DAPNIA/SAp, 91191 Gif-sur-Yvette Cedex, France
}

\received{3 Dec 2007}
\accepted{14 Dec 2007}
\publonline{13 May 2008}


\abstract{%
This article quickly summarizes the performances and results of the
GOLF/SoHO resonant spectrometer, thus justifying to go a step further.
We then recall the characteristics of the multichannel resonant
GOLF-NG spectrometer and present the first successful performances of
the laboratory tests on the prototype and also the limitations of this
first technological instrument. Scientific questions and an observation
strategy are discussed.}

\maketitle

\section{Introduction}
The SoHO satellite first milestone of the ESA Horizon 2000 programme, 
is a real success due to the simultaneous presence of internal and external
solar probes. It enriches the present transition period between a classical view
of stars and a dynamical view that we are beginning to constrain. The
success is partly due to helioseismology which reveals the properties of
millions of modes not yet within reach for any other type of
stars. This fact enables a strong scientific return with some problems
being solved and totally new questions emerging.

Seismology of the Sun and stars will be pursued during the next decade
with a strong impact in Astronomy.  The successors of SoHO are SDO in the United
States (launch in 2009) and Solar Orbiter in Europe (launch in 2014-2015). But
these missions have not been designed to probe the deep solar interior. 

After more than a solar cycle of continuous observation with SoHO and three
decades with the BiSON ground network, it is time to critically review
the results obtained by the European community to highlight the
advantages and the limitations of the resonant spectrometer instrument
and to deduce which directions will lead to
improvements.  In section 2, we recall the characteristics of the GOLF
instrument and its results. The third section deals with the objectives
of the GOLF-NG instrument, successor of GOLF,  the performances obtained in
the laboratory and the necessary improvements to get a scientifically
relevant instrument.  The last section will be devoted to (1) the questions
that we need to solve in the future,   (2) the solar proposal at the ESA Cosmic
Vision level and (3) the way we might observe during the next decade. 

\section{Probing the nuclear core}
Three instruments from the SoHO satellite (GOLF, VIRGO and
MDI, Domingo 1995) have now been observing the Sun down to the core for
more than ten years and two ground networks (GONG and BiSON) have accompanied 
this effort, as they have been operating for the last two or three decades. This
redundancy allows us to compare the quality of the observations and to study
 possible progress. We note that GOLF has been the most successful for
the exploration of the solar nuclear core which contains practically half the
solar mass and for which the dynamics was totally ignored before the SoHO
launch. Moreover, despite the observing conditions of BiSON which are more
difficult than around the L1 Lagrangian point, such a network is extremely
useful and reaches very good performances for the low degree acoustic modes due
to the fact that we can integrate the information for a long period of time
(Brookes et al. 1978). In the two cases resonant spectrometers are used.
\subsection{Comparison of the techniques used aboard SoHO} 
GOLF measures the variability of the Doppler velocity by the resonant
spectrometer technique on  three sodium lines: D1, D2a and D2b (Gabriel
et al. 1995). The name of this instrument, Global Oscillations at Low
Frequency, was given because it has been designed to measure very small
velocities (down to 1mm/s) in order to get the low-degree low-frequency
modes, \textit{i.e.} the modes which penetrate deeply in the radiative
zone down to the core.  The method has been invented by Brookes, Isaak and van
der Raay (1978) and is used  in the IRIS and BiSON
ground networks. The GOLF instrument filters the solar light at wavelengths
corresponding to the mentioned lines. The corresponding photons, left or right
polarised, are absorbed by the atoms of a hot sodium cell and re-emitted in a
narrow band, shifted thanks to the presence of a permanent magnet which produces
a Zeeman effect. The Doppler velocity  is theoretically deduced from the
comparison of the left and right circularly polarized photons. Its time
variation is measured every 10s. The power spectrum obtained by Fourier
transform of the velocity signal is shown in Figures 1 and 2. Despite the
malfunction of the quarter wave plate motor, this measurement has been done on
one wing (left or right depending on the period of observation) during the 12
years of the SoHO mission (Garc\'ia et al. 2005) thanks to the presence of a
small modulation of the magnetic field.  Several comparisons have been made
between  GOLF performances in this configuration and the BiSON and GONG
networks and also between GOLF, VIRGO and  MDI.  
\begin{figure}[h]
 \small  
  \centering
   \includegraphics[width=8cm]{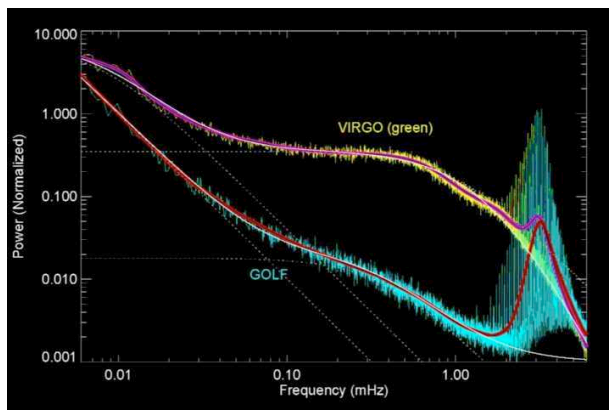}
    \caption{Comparison of the power spectrum of GOLF and VIRGO which look
             at the Sun as a star from SoHO. GOLF uses the
             Doppler velocity technique and VIRGO the variability of the
             luminosity at different wavelengths.  The first technique is
             largely better at low frequency where acoustic modes are not
             perturbed by the solar cycle (below 1.6 mHz) and where 
             gravity modes are expected (below 0.4 mHz).  From Bedding and
             Kjeldsen (2006).}
    \label{myfig1}
 \normalsize         
\end{figure} 

Figure 1 shows the superiority of the Doppler velocity technique over
measurements of  luminosity variations through two
instruments located aboard SoHO.  During the first two years of
observation, one could notice a difference of up to a factor 10 to 30 in
the Fourier transform at low frequencies in the range of 
gravity modes located between 10 to 400 $\mu$Hz. This is due to two factors.
First the GOLF velocity is extracted from the solar atmosphere at a height of
300 to 500 km and this region is less turbulent than the photosphere. Secondly,
the GOLF instrument has been specifically designed to detect very low amplitude
signal at low frequency down to 1 mm/s in the range of the first gravity modes.
This performance is obtained by using photomultipliers and a counting rate of
$1.2 \times 10^7$ photons/s associated to  very stable
electronics. Consequently the instrumental noise was significantly smaller than
the statistical noise which was chosen as low as possible in relative
value (at least during the first years). So, this noise contribution was
reasonably flat at low frequencies in the range of gravity modes without any
atmospheric perturbation (see Figure 2).
\begin{figure}[h]  
   \centering
      \includegraphics[width=7.5cm]{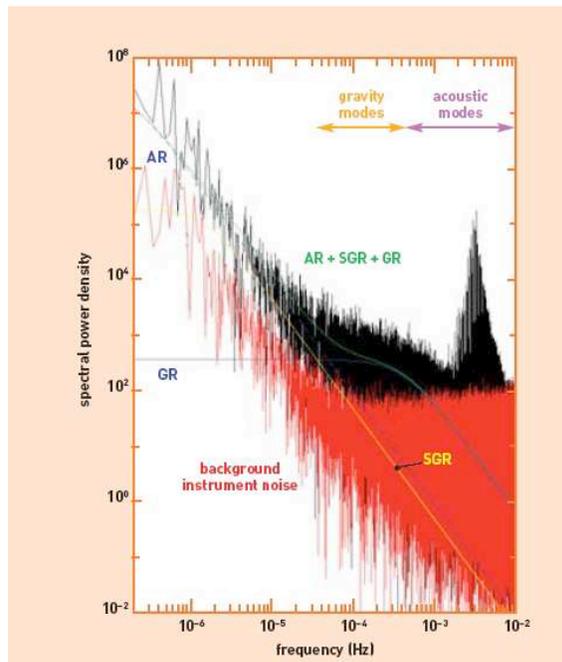}
      \caption{GOLF velocity measurement power spectrum. We have
               shown explicitly where acoustic and gravity modes
               are located
               together with the different contributions which form the solar
               noise. The instrumental noise is superimposed in red. From
               Turck-Chi\`eze et al. (2004a).}
         \label{myfig1}
\end{figure}
\subsection{Scientific results}
The GOLF instrument has been successful in answering the different questions for
which it has been designed. There are three domains where it has given very
positive answers:  
\begin{itemize}
\item the description of the solar core and the prediction of the emitted
      neutrinos thanks to the detection of previously unknown low-degree
      low-frequency acoustic modes not polluted by the variability of the
      external layers during the solar cycle. The sound speed is
      precisely determined down to 6\% R$_\odot$ and the deduced neutrinos in
      agreement with detections (Turck-Chi\`eze et al. 2001, 2004b,
      Turck-Chi\`eze \& Talon 2008). 
\item a better understanding of the impact of the solar cycle on
       subsurface layers by  low degree acoustic mode
      variations above 1.6 mHz (Gelly et al. 2002,  Garc\'ia et al. 2008a).  The
      evolution of these modes during the Hale 11 year solar cycle
      also benefit from the three decades of observations
      with the BiSON network (Chaplin et al.  2007). They allow
      us to put some constraints on the local and varying
      subsurface magnetic field (Nghiem et al. 2006). 
\item  the ability to detect gravity modes. We have
      mentioned that the GOLF instrument has been specifically designed
       to
      detect some gravity modes. But the publication of Kumar et al.
      (1996), after the SoHO launch, on very low surface
      velocities coupled to the malfunction of the polarizers have put some
      doubts as to GOLF's ability to detect any signal. Nevertheless, a
      search strategy (among others) has been defined:  first to look for
      multiplets (it improves the ability to detect gravity
      mode signal) in the region above 150 $\mu$Hz where  velocities
      are  greatest and then to examine the low frequency region if the
      first search was successful. In both cases, signals have been
      detected with more than  a 98\% of
      confidence level (Turck-Chi\`eze et al. 2004a, Turck-Chi\`eze 2006, Mathur
      et al. 2007) and even at low frequency at more than a 99.7\% of
      confidence level (Garc\'ia et al. 2007). 
\end{itemize}
\subsection{Limitations of  present observations}
We are today confident that GOLF has detected the signature of dipolar solar
gravity modes. Nevertheless the information on the rotation profile in the
nuclear core remains rather poor due to some uncertainties on the
identification of the observed components or to the fact that it is difficult to
extract the splitting of the modes located in the lower part of the spectrum
(Garc\'ia et al. 2007, Mathur et al. 2008).  It is interesting to note that the
two analyses seem compatible only if the core of the Sun turns quicker than the
rest of the radiative zone and is an oblique rotator; this idea was already
discussed a long time ago (Isaak 1982). Clearly  this conclusion needs to be
confirmed by improved observations because the scientific consequences of such
a behaviour have a high level of interest. Presently, the analysis of the other
instruments (MDI, VIRGO, BiSON and GONG networks) has shown the superiority of
space instruments in the gravity mode range: one component of the GOLF gravity
mode candidate  appears continuously in the VIRGO data (Garc\'ia et al. 2008b).
Unfortunately the BiSON data has not yet revealed any signal in this range
(Broomhall et al. 2007). 
 
Considering  present observing conditions of GOLF, we know since 1998
that we need to go further. The CNES microsatellite PICARD mission selected at
that time, will be launched in 2009. It will  explore the potential increase of
 gravity mode sensitivity at the solar limb. An enhancement by a factor of 4
or 5 has been measured with MDI in the range of acoustic modes but it is not
evident that such an effect would be sufficient to detect gravity modes
with a photometric instrument when considering Figure 1. So it is
natural to also push the development of a new instrument
using the Doppler velocity technique which derives from our expertise in Europe.
The first pollution to limit a good signal/noise ratio in the gravity mode range
comes from the Sun itself, so the idea for decreasing this noise is to measure
the velocity at different heights in the atmosphere which do not experience the
same granulation noise (Espagnet et al. 1995). Trying to do so supposes the
exploration of a line sufficiently broad to examine the total range between the
photosphere and the chromosphere (see   Simoniello et al.
2008, this volume). In that sense, the D1 sodium line is
well justified for this purpose. This is the reason why we are pushing
the performances of a sodium resonant spectrometer to its limits. The GOLF-NG
instrument will be described in the next section. The main objective of
this instrument is to reduce both the solar noise and the instrumental noise by
a factor of 5 to 10. 

\section{The GOLF-NG prototype}
\subsection{A multichannel resonant scattering spectrometer}
Like GOLF, the GOLF-NG (Global Oscillation at Low Frequency New
Generation) instrument measures  global Doppler velocity variations. It
is being developed in the CEA/France in collaboration with
the IAC/Spain. It results from   30 years of expertise
on resonant scattering spectrometers used on the ground (IRIS and BiSON
networks) and in space (GOLF/ SoHO).  The characteristics of this instrument are
described in Turck-Chi\`eze et al. (2006). The objectives in space (or possibly
on the ground) are to lower the mode detection threshold by about a
factor of 5, to detect easier acoustic modes,  to identify
different components of several individual gravity modes and to pursue the
analysis of the asymptotic behaviour of the modes at low frequency for at least
$\ell$ = 1, 2, 3 and maybe 4 and 5. The improvements are the following:
\begin{itemize}
\item we measure simultaneously the velocity at 8 positions between the
      photosphere and the chromosphere to reduce the noise due to 
      solar granulation in the range of g-modes. As the granulation patterns
      change with altitude (Espagnet et al. 1995),  we will benefit from the
      lack of coherence in the solar noise to improve the signal
      to noise ratio in the range of gravity modes; this
      point has been already shown with GOLF data (Garc\'ia 2004),
\item we increase the number of photons detected to also reduce the relative
      instrumental noise. This relative noise increases with time in the GOLF
      instrument so we will try to avoid degradation with time to have long
      observation periods in the best conditions. For this purpose, we have
      multiplied by 2 the number of detectors per height in the atmosphere: 4
      detectors instead of 2, and we must use a stable detector with a higher
      quantum efficiency (60-75 \% instead 5\%). Altogether we get 32 outputs
      (in fact 31 for mechanical reasons) from the cell (instead of 2 in GOLF),
\item the gravity mode spectrum is very dense so the identification of
      components in the observed pattern requires the use of
      some masks at the entrance of the instrument. Detection and identification
      of degrees up to 5 for gravity modes is an objective for the coming
      decade and might allow a precise determination of the rotation profile in
      the whole solar core,
\item adding an entrance polarizer could help measure the mean
      magnetic field and its time evolution, like in the nominal operation of
      GOLF. It has been used in space during the first month (Gabriel et al.
      1995, Garc\'ia et al. 1999, Chaplin et al. 2003). 
\end{itemize}
This instrument measures the Doppler shift of the D1 so\-dium solar line alone
(in contrast to GOLF which was using the 3 lines D1, D2a, D2b only
knowing their mean position). The associated photons
are compared to those of an absolute standard given by the sodium vapour cell,
the heart of the instrument. A small portion of the line is measured by the
resonant photons  which escape from the vapour cell. It is split into its Zeeman
components by means of a longitudinal magnetic field, the strength of which
varies linearly along the axis of the magnet to explore different heights
in the atmosphere located between 300 and 1000 km (Jim\'enez-Reyes et
al. 2007). One selects simultaneously 8 points on the right wing of the line or
8 points on the left wing, including one fixed point at the center of the line
(supposing no shift of the line, see Figure 3) by changing the circular
polarization of the incoming flux thanks to liquid crystals. It  avoids the
change of polarisation by a motor and it reduces the weight of this instrumental
part. The instrument must measure a  flux high enough to reduce the instrumental
noise or (and) to allow consecutive measurements of portions of the Sun with a
good statistics and no saturation. A second liquid crystals polariser could be
installed at the entrance to get a spectrum of the mean magnetic field. 

A second objective of this instrument is to put some constraints on the
influence of the magnetic field on the solar atmosphere. In measuring
the sodium line in 15 points every 2 seconds, we will probably be able to
check  atmospheric models and their evolution with  solar
activity.
 \begin{figure}[h]  
   \centering
    \includegraphics[width=7cm]{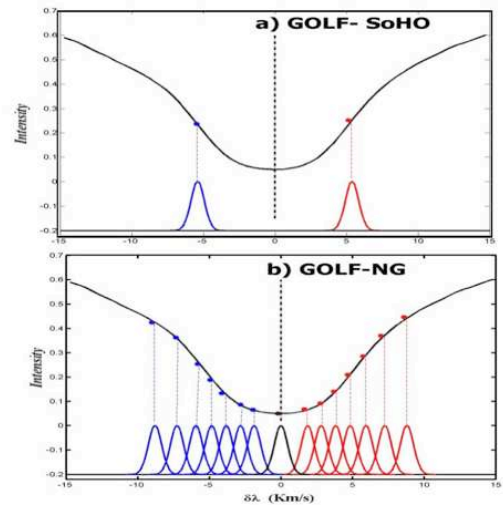}\hspace{4pc}
      \caption{Solar resonant spectrometer  GOLF-NG principle: an alternating
               measurement of 8 points along the blue wing and then 8 points
               along the red wing ensures a measurement of 15 points along the
               line in 2s (or less) to extract the velocity at different heights
               in the atmosphere. These different points are obtained by putting
               the cell inside a permanent magnet  delivering a magnetic field
               varying quasi linearly from 2 to 8 kG.}
         \label{myfig1}
 \end{figure}
\begin{figure}[h]  
   \centering
   \includegraphics[width=5.5cm]{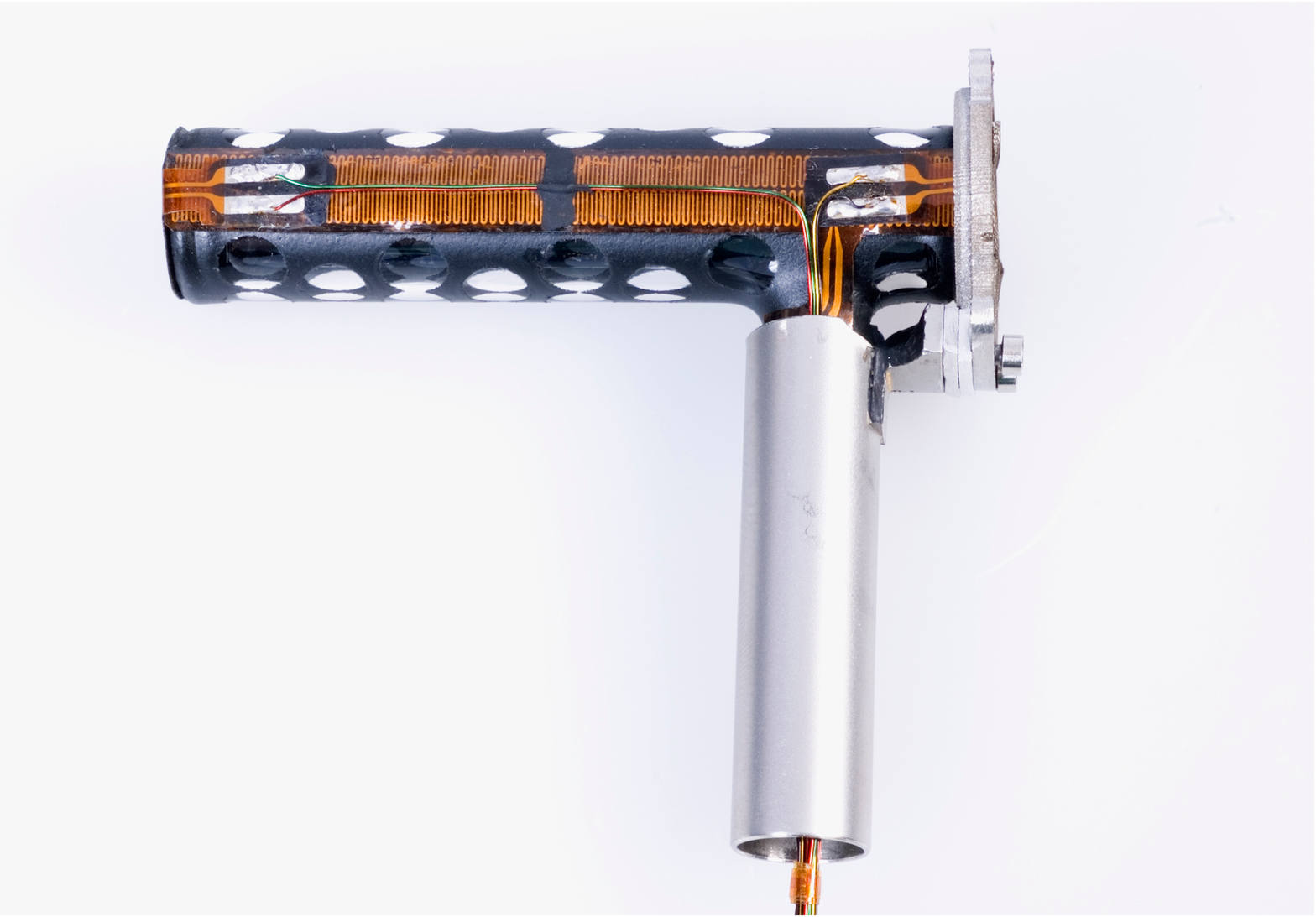} \vspace{4pc} \includegraphics[width=5.4cm]{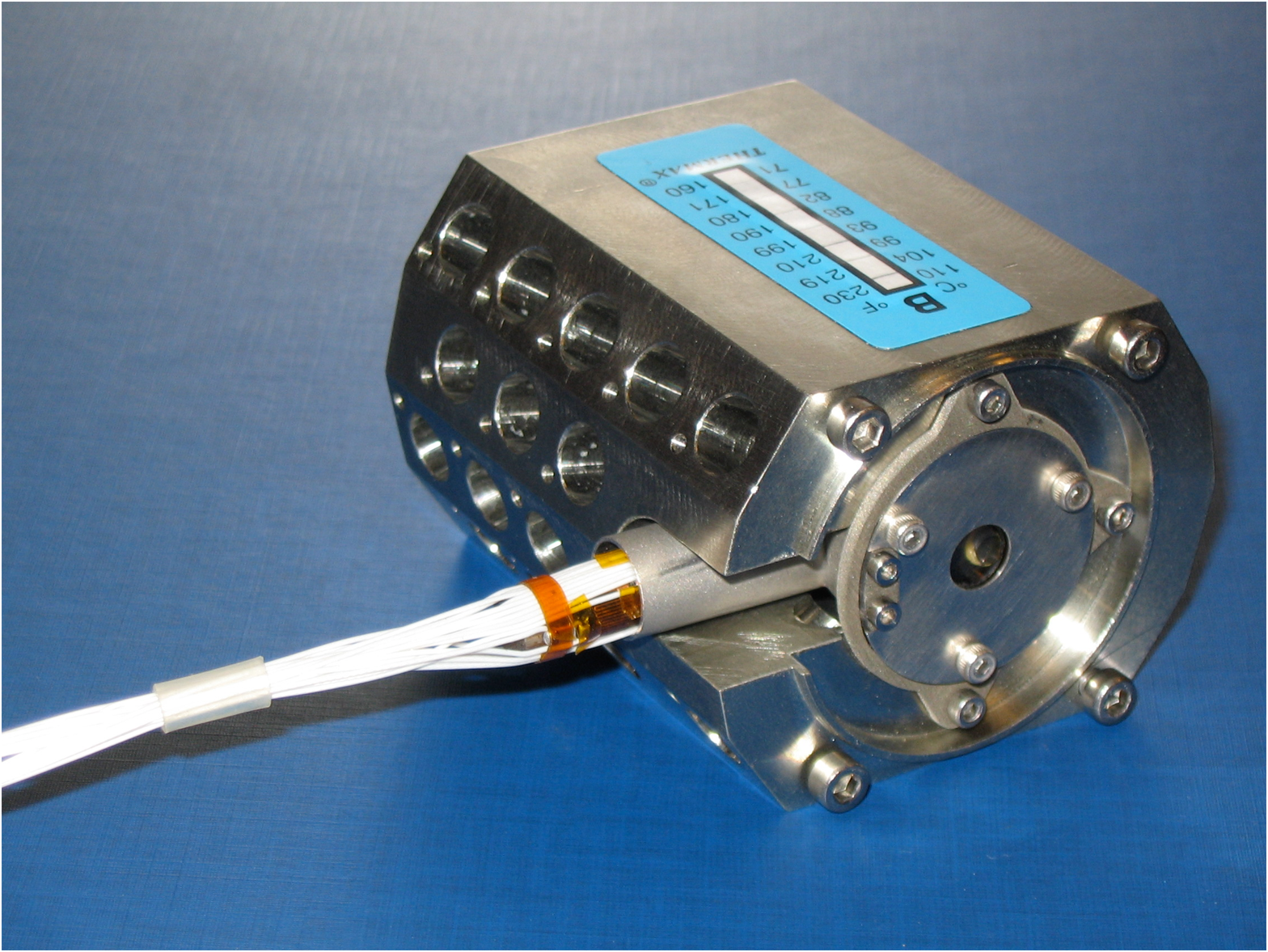}  \vspace{2pc}
 \includegraphics[width=7cm]{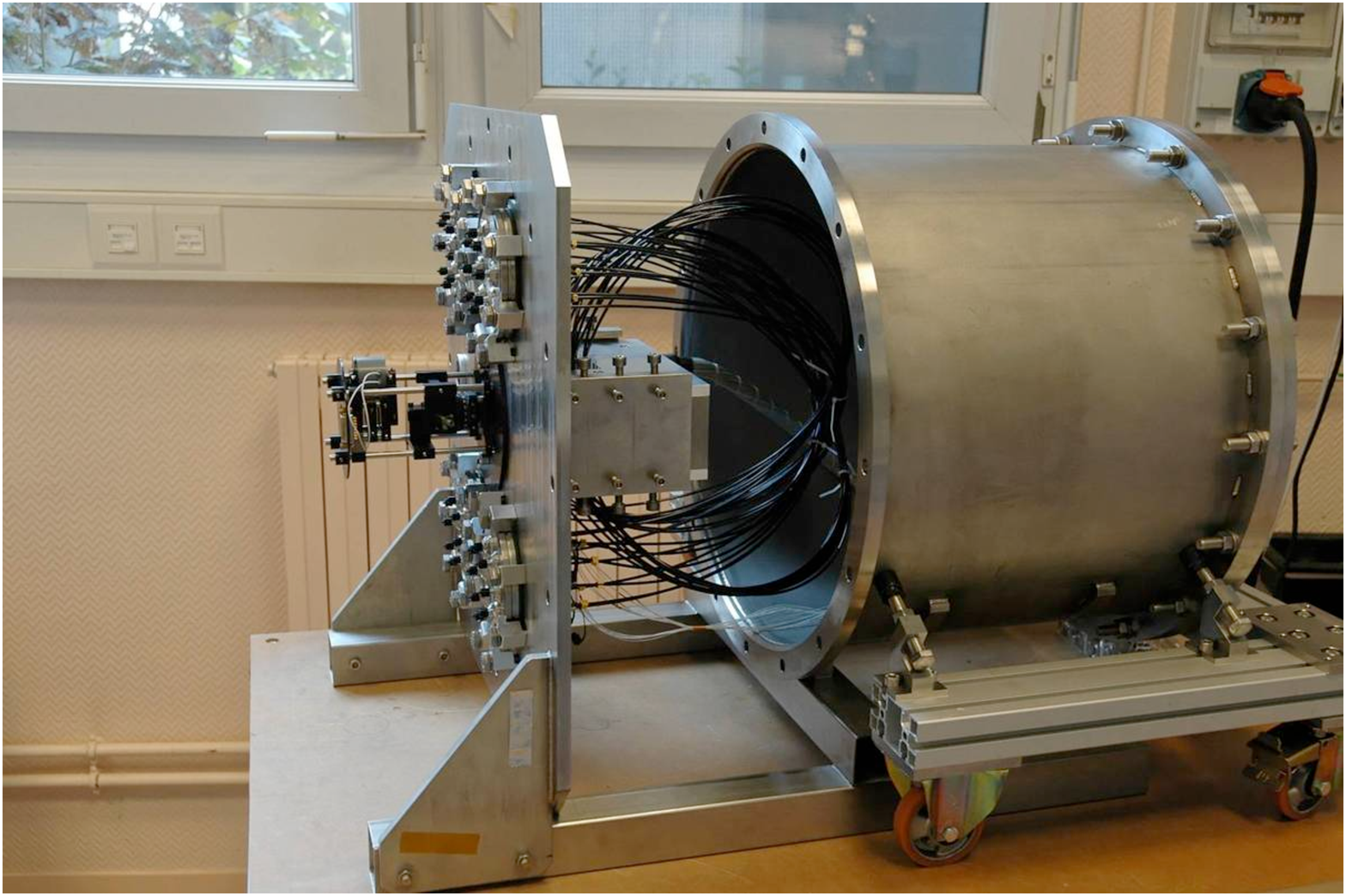}  
 \vspace{-5mm}  
      \caption{a) View  of the long cell equipped with the heaters and prepared
               with its 31 outputs (8*4-1). b) The same cell inside an insert
               piece designed in Saclay and constructed at the
               IAC. It ensures the cell's mechanical position
               and the thermal difference between the cell's high
               temperature and the low temperature of the magnet. c) A
               view of the GOLF-NG prototype functioning in a vacuum to
               respect the thermal conditions in space.}
         \label{myfig1}
 \end{figure} 
 \vspace{-0.4cm}
\subsection{The subsystem level}
During the last 3 years, time has been devoted to solving the
critical technical problems (Carton et al. 2008) we need to face to
succeed  measuring the 31 resonances along the cell
(Figure 4). The sub-systems have been studied separately to estimate the
performances of the prototype (magnetic field linearity, response of the
photodiode matrix detector, dark current, optical design, thermal
equilibrium of the cell, reflection of light ...). GOLF-NG is an
extremely complex instrument to construct because it needs
to achieve very good performances to be able to detect signals
corresponding to
velocities as small as a fraction of mm/s. It needs (1) a permanent
magnet of small size varying linearly between 0 and 12 kG, (2) good thermal
conditions of the cell heated presently to around 170$\dg$C and located
inside a magnet maintained at a temperature around 25$\dg$C thanks to (3) an
heavy insert piece (see Figure 4). The long but small cell (8 mm* 60 mm) is
filled with pure sodium and must be used with caution to prevent any
deterioration of the glass properties and (4) to keep a good thermal equilibrium
of the cell bulk despite the 31 outputs  with a temperature homogeneity within
only several $\dg$C and a stem heated to a temperature lower by
10-15 degrees. The difficulties are to limit the reflection of  light
in the cell, a  loss of counting rate between the cell and the detector. (5) The
chosen detector for the prototype is able to measure a high flux (up to 1.5
10$^8$ph/s) per photodiode without saturation. Unfortunately,
the electronic noise is too high (see Mathur 2007).  In fact this electronic
noise needs to be about 1/10 of the statistical uncertainty to get an
instrumental  white noise in the range of gravity modes instead of an
instrumental noise increase below $10^{-3}$ Hz. 

 \begin{figure}
\includegraphics[width=80mm,height=80mm, angle=90]{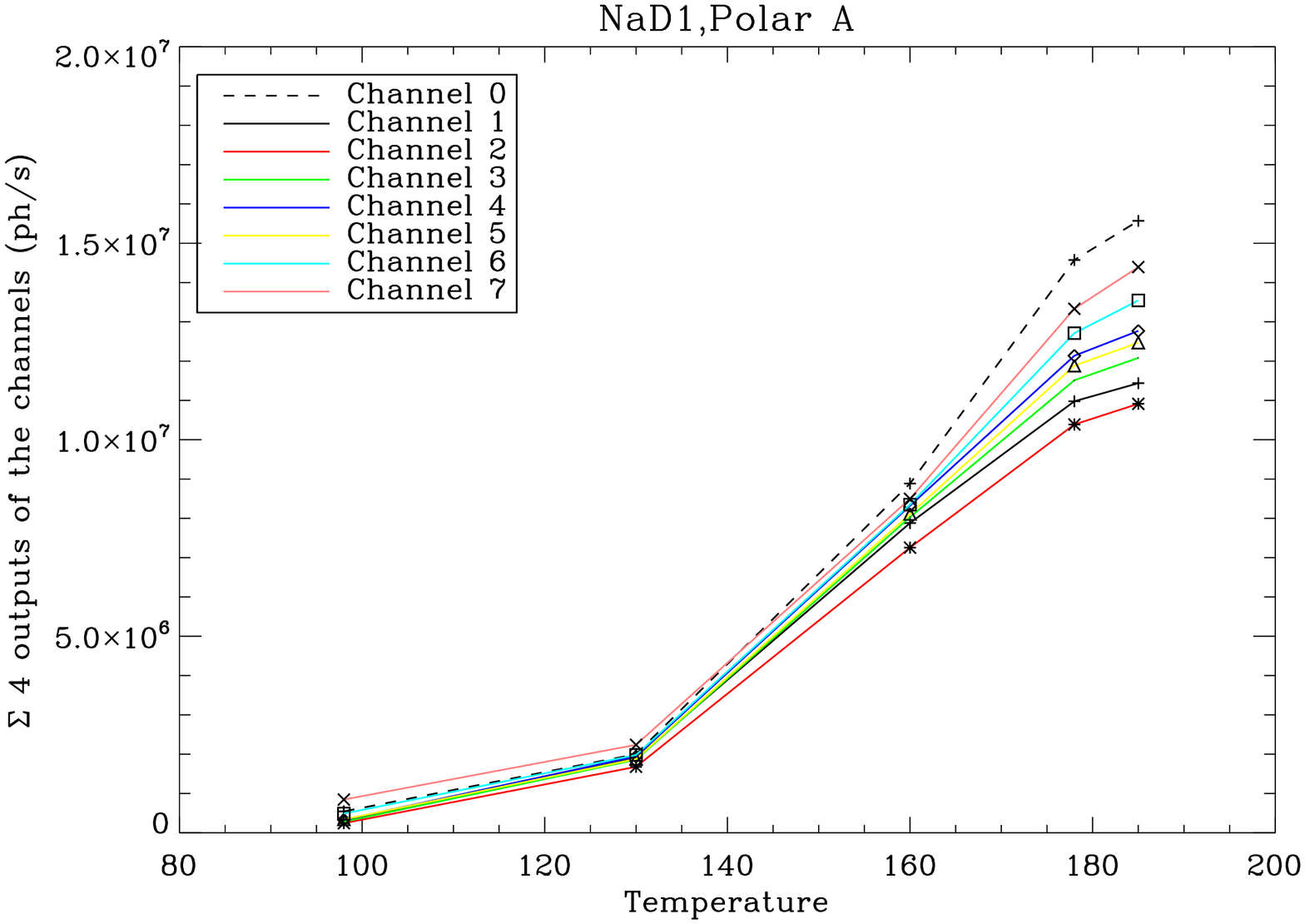}
\caption{Mean value of the resonant photon response of each channel as a
function of the temperature of the cell using a calibrated LED at the entrance
of the instrument.}
\label{label1}
\end{figure}

\subsection{Laboratory tests}
Most of the hard points have been solved at the sub-system level and a complete
prototype equipped with Hamamatsu photodiodes has been studied during the year
2007. 

The laboratory tests are done in space conditions. It is useful to study
the critical thermal conditions i.e. in a vacuum tank (see Figure 4c) to prepare
this instrument for space measurements. 
\begin{figure*}
\centering
\includegraphics[width=140mm,height=180mm]{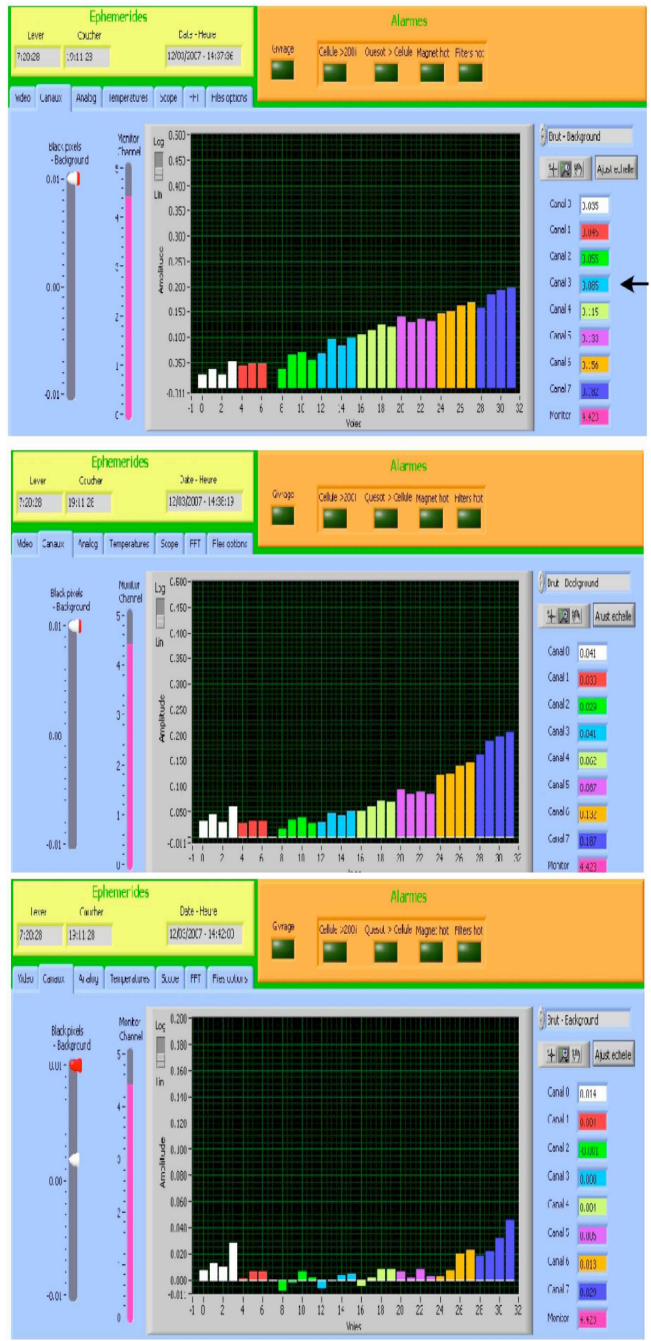}
\caption{A quick look at  solar measurements done in
March 2007 at Saclay during half an hour. One observes the resonance on the 31
channels for the NaD1 filter in the two polarity measurements (left and right of
the line on the first two figures) and the absence of resonance (last figure)
when the second filter is chosen (corresponding to a continuum region where we
have no sodium line), in which case we measure the scattered light.  In
the different measurements, the subtraction of the detector background is
already applied.}
\label{label1}
\end{figure*}

Different tests, which we briefly summarize here, have been
performed, first with a LED and then with the Sun. At
the entrance  of the instrument, two filters can be used: the first one is
located around the NaD1 line ($\sim$589.6 nm) with a width of 0.6 nm, the second
one is centered on the continuum ($\sim$591 nm) with a  larger width. These
filters enable us to distinguish the luminosity variation (if we look directly
at the corresponding photons without looking through the cell) from the
velocity variation (measurement of the photons absorbed and re-emitted by the
sodium cell) during  solar observations. When we use a lamp at the
entrance of the instrument or when we look at the Sun, the continuum measurement
through the cell allows us to measure  scattered light inside
the cell. 

The resonance has been obtained for each channel with a LED and Figure 5
shows how the resonance flux evolves with the temperature. One notes that the
behaviour is similar for the 8 positions and that the dispersion between
the channels, which is not so large and results from the
dispersion between the photodiodes, can be corrected.

A rapid measurement with the Sun has been performed in March 2007. Figure 6
shows that the resonance appears clearly for the two polarizations and the shape
of the photon flux along the cell corresponds reasonably well with what is
expected from the different points along the line. Complementary
measurements have also been done with a laser to properly estimate the
wavelength width of the different channels (typically 25 mAngstrom) illustrated
in Figure 3. 

Several corrections are taken into account: the dark current including the
electronic noise and the reflection light (measured at low temperature) or the
scattering light measured with the continuum option. These corrections  seem
much more severe for the extreme channels with the Sun, in particular
for the channels at the bottom of the line, and must be improved
in the scientific instrument. 

We have deduced from the first series of measurements the response of the whole
system. The resonant flux was smaller than expected by
about a factor 4. Complementary measurements have been done during the year and
a new campaign of measurements with the Sun in Saclay has been done in March
2008.  The resonant flux is presently nearly nominal.
A campaign of tests is scheduled at Tenerife to study the interest of this
multichannel resonant scattering spectrometer, even though the nominal
performances have not yet been reached with this prototype and only  allow a first estimate of the measurement 
of the solar noise at different heights in the atmosphere.

\subsection{Towards a scientific instrument}
Several problems must be solved before producing  an instrument scientifically
representative of a new generation resonant spectrometer for space
observations: 
\begin{itemize}
\item we cannot avoid having the sodium vapour attack the pyrex
      cell at high temperature (the GOLF instrument was equipped with a
      galeniet glass). This problem limits the resonance temperature and
      consequently the lifetime of the cell. The prototype must be used  at
      a lower temperature than the nominal one (165 instead  of 190
      degrees) with a reduction of the flux by practically a factor of 2. So we
      are investigating manufacturing another cell for the scientific
      instrument to avoid this problem which is only a limitation in the final
      flux (which may also be compensated by other actions). We
      will also estimate the cell lifetime with pyrex at a
      reasonable temperature.
\item we need to use a detector which delivers 2.5$\times$10$^7$ph/s per channel
      at mid height of the line  with an intrinsic noise significantly smaller
      than the statistical one. It is not the case today for 
      arrays of Hamamatsu photodiodes containing an integrated
      electronic which produces an electronic noise at the same level as the
      statistical noise for an optimal temperature of 6 degrees. Such
      limitations have a direct impact on the behavior of the
      instrument at low frequency leading to an increase of the instrument
      background  below 1 mHz which competes with the solar background.  We hope
      to use a CCD detector and add together the pixels corresponding to
      a spot of about 1.3*1.3 mm$^2$. This detector will be cooled to
      the appropriate temperature (-40 or -50 degrees) to get a low instrumental
      noise largely dominated by the statistical noise (by about a factor
      of 10),
\item for an optimal response of the whole instrument, we could slightly
      increase the entrance pupil, the response of the filters or the quantum
      efficiency of the detector.  We will estimate this need after the
      determination of the flux response of the prototype to  sunlight
      in the laboratory (partly done already) and in Tenerife,
\item we need to study some masks at the entrance of the instrument. This will
      be useful in order to take advantage of the amplification of the signal at
      the limb or to identify without ambiguity the observed patterns when the
      first g-modes will appear.
\end{itemize}

So after the first tests in a helioseismic station, the next step will be
the building of a scientific instrument hopefully more compact and
having improved performances that we could put in different sites. It seems
today that Tenerife is a good site to analyze the behavior of the solar noise at
different levels in the acoustic range  and then in the gravity mode range when
the detector will be sufficiently good to avoid any instrumental
perturbation. Then a measurement in Dome C could be determinant to try
to observe g-modes on ground during 1 or 2 campaigns of several months. 
If such a step demonstrates the potential of the GOLF-NG
concept, it will be easy to extrapolate it to a space version very
quickly. We note that the development of an operational multichannel resonant
scattering spectrometer needs a very dedicated effort during several years if we
want to detect properly gravity modes in a reasonable period of time (at least
one or two months for real constraints on the rotational splittings and then
several months for some magnetic field displacement of these splittings). 

\section{Perspective for the next decade}
The success of SoHO  invites the European community to pursue its investigation 
of the Sun as a whole because it is the only star which can deliver an
internal information on all the processes in action in stars. SoHO and CLUSTER
have also revealed a strong interaction between the Sun and the
Earth. Knowing the different origins of  solar activity justifies a
continuous and permanent observation of our star from the core to the corona.
SDO and Solar Orbiter will only partly cover this objective. This is the reason 
why we have described new scientific perspectives during the Cosmic Vision
meeting at ESA  in 2005 and prepared the DynaMICCS formation flying mission
which is one way to answer the questions through a dedicated unique mission
(Turck-Chi\`eze et al. 2005, 2006, 2008). It is interesting to note that the
origins of the Maunder minimum or of the historical maxima must be connected to
the understanding of the dynamical processes of the solar machine including the
dynamics of the radiative zone. We need such information for a proper prediction
of solar variabilities during the next century. 
\subsection{The new scientific objectives}
A lot of questions remain unsolved and will not be solved by the already
confirmed missions. They correspond to the knowledge of the dynamics of
the radiative zone which represents 98\%  M$_{\odot}$ and more specifically of
the core with more than 50\%  M$_{\odot}$ and require gravity mode detections
and associated theoretical efforts. They can be summarized here: 
\begin{itemize} 
 \item What is the dynamical influence of the central rotation and of the
       magnetic field on  external activity? 
 \item Which processes are at the origin of the solid body rotation observed in
       the radiative zone down to 0.2 solar radius?  What is the respective role
       of the agents responsible for the redistribution of the angular momentum:
       rotation, gravity waves, magnetic field?  What are the consequences of a
       rapidly rotating core?  Is there another dynamo in the core?
 \item What is the topology, strength and influence of a fossil field?  How
       do progressive internal waves modify the overall internal
       dynamics?  Could we determine the nature of the nonlinear interactions
       between the convective dynamo and the fossil field if it exists?
 \item How do we check the presence of large scale flows, their amplitude and
       their mixing properties in the radiative zone?  Could we put some
       constraints on the presence of magnetohydrodynamical instabilities in the
       radiative zone and their coupling with the convection zone?
 \end{itemize} 
These questions show the need to understand  deep solar magnetism and deep
solar motions in their different forms. 

The  mission called DynaMICCS for { \it Dynamics and Magnetism from the Core to
the Corona of the Sun}  is conceived to measure the observables which must
identify the different sources of the solar cyclic variabilities. To reach this
objective, crucial regions of the Sun must be scrutinized simultaneously: (1)
the previously unexplored dynamics of the inner core thanks to gravity modes,
(2) the time evolution of the radiative/convective zone interface layer thanks
to a large number of acoustic modes, (3) the emergence of the flows from the
photosphere to the chromosphere layers thanks to the study of different lines
and different heights in the atmosphere, (4) the evolution of the low corona
never explored continuously thanks to a permanent eclipse, (5) the total and
spectral irradiance and (6) in-situ measurements of plasma/energetic
particles/magnetic fields of the solar wind. 

The complementary instruments are  well identified but unfortunately this
mission will not be scheduled in ESA before 2020 due to the delay of the Solar
Orbiter mission.

Among them, GOLF-NG has a crucial place if its performances are in accordance
with its objectives.

\subsection{Progressing on the solar core dynamics during the next decade}
It seems very important to estimate the potential of the new technique
of GOLF-NG in comparison with GOLF, the BiSON network or the
SODISM instrument aboard PICARD. While building the Solar Dynamical Model (SDM)
instead of the Standard Solar Model (SSM) (Turck-Chi\`eze 2008) and
developing more and more realistic 3D solar simulations (Brun \& Zahn 2006), it
is important to find a way to constrain these models with a scientific
instrument like GOLF-NG during the coming decade, to add constraints on
the solar core dynamics and to  explore 
convective dynamics with SDO and PICARD.

We hope to accomplish the following steps:   first, a campaign with the
prototype in Tenerife to see how the acoustic modes behave at different heights,
and what we learn from the line and the atmosphere, if the present
detector allows these measurements; then we hope to develop a scientific
instrument for GOLF-NG that would have the characteristics of a space
version, in terms of weight and size. With this instrument, we will estimate the
interest of this new multichannel spectrometer at very low frequency (the
gravity-mode range) in comparison with the instruments of the BiSON network, HMI
aboard SDO and SODISM aboard PICARD.  We would like to check its performances at
Dome C through one or two rather long and continuous observation
campaigns  of at least 1 month to be able to really quantify the
performances and see if one needs to improve them. The polar location
and the data continuity of Dome C would be crucial for validating the
performances and we hope to detect the already seen g-modes.  The important
periods for the observations of GOLF-NG will be the years 2009-2012 where SoHO
observations will be finished or difficult with GOLF. During this period, PICARD
and SDO will be operational. The schedule is tight and assumes a good
organization. 

If the performances of GOLF-NG are satisfactory, it will be integrated in a
space mission as soon as possible.  The best would be to overlap
in time with the SDO mission. GOLF-NG could be also included in a
ground network.

Helioseismology in Europe constitutes a long and dedicated effort for the
last three decades, in which it is at the origin of a lot of
discoveries and the transition to a multiscale stellar physics. The
present project benefits from a French culture developed in the IAS,
Nice, Saclay and Meudon  with some interaction with Themis. Moreover, the
expertise of British and Spanish teams are important. Thus, such effort must be
pursued at low cost for at least one decade continuously. 

This discipline largely contributes to the understanding of the dynamics of
stars and will play, in the future, an important role on the long term Sun-Earth
relationship. It will enrich the development of  dynamical stellar evolution.
Finally, it will indirectly contribute to the knowledge of how planets evolve
with their stellar environment, a new direction of very promising improvements
with KEPLER, GAIA and PLATO.
 
\section*{Acknowledgments}
The GOLF-NG instrument is developed under the responsibility of S. Turck-Chi\`eze
and P.H. Carton by an international consortium joining the CEA/Saclay team and
the IAC team directed by P. Pall\' e in Spain. The whole consortium will
publish detailed analysis of the GOLF-NG prototype and wishes to
thank the GOLF and IRIS teams for transmitting  their useful
expertise for this instrument.  The DynaMICCS mission proposed to ESA has
evolved through the DynaMICCS/HIRISE consortium. This collaboration is supported
by scientists coming from 30 institutes from Europe, India and
the United States.


\begin{thebibliography}{}
\bibitem{} Bedding T. R.  and Kjeldsen H.: 2006, {\it Mem. Della Soc. Astron. Ital.}   {\bf 77},  384
\bibitem{Bertello2000} Bertello L., Henney C. J. and Ulrich R. K.:  2000, {\it ApJ} {\bf 537},  L143
\bibitem{} Broomhall, A. M., Chaplin, W.J., Elsworth, Y.: 2007, {\it MNRAS}  {\bf 379}, 2
\bibitem{} Brookes J. R., Isaak G. R. and  van der Raay H. B. :1978,  {\it MNRAS}   {\bf185},  1-17
\bibitem{} Brun, A. S. \& Zahn, J.P.: 2006, {\it ApJ} {\bf 457} , 665 
 \bibitem{}Carton P. H., Turck-Chi\`eze S et al.: 2008,  {\it to appear in Experimental Astronomy} 
\bibitem{} Chaplin, W. J., Dumbill, A. M., Elsworth, Y. et al.: 2003, {\it MNRAS}  {\bf 343},  813
\bibitem{} Chaplin, W. J., Elsworth, Y., Miller, B.A. et al.: 2007,  {\it ApJ} {\bf 659}, 1749
\bibitem{}Domingo V., Fleck B. and Poland A. I.: 1995, {\it Sol. Phys.}  {\bf 162}, 1
\bibitem{}Espagnet O., Muller R., Roudier, Th., Mein, N., Mein, P.: 1995, {\it A\&A Suppl. Ser.}  {\bf 109}, 79
\bibitem{} Gabriel A., Grec G., Charra J. et al.: 1995, {\it Sol. Phys.}  {\bf 162},  61
\bibitem{} Garc\'ia R. A., Boumier P., Charra, J. et al.: 1999, {\it A\&A}  {\bf 346}, 626
\bibitem{Garcia01}Garc\'ia R. A., Regulo C., Turck-Chi\`eze S. et al.: 2001,  {\it Sol. Phys.}  {\bf 200},  361
\bibitem{} Garc\'ia R. A., Jim\'enez-Reyes S. J., Turck-Chi\`eze S., Mathur S.: 2004, ESA-SP  {\bf 559}, 432
\bibitem{} Garc\'ia R. A., Turck-Chi\`eze S., Boumier P. et al.:  2005, {\it A\&A}  {\bf 442},  385 
\bibitem{Garcia07}Garc\'ia R. A., Turck-Chi\`eze S., Jim\'enez-Reyes S. et al.: 2007, 
{\it Science}  {\bf 316},  1537
\bibitem{Garcia08}Garc\'ia R. A., Mathur S., Ballot J. et al.: 2008a, {\it Sol Phys} in press 
\bibitem{Garcia08}Garc\'ia R. A.,  Jim\'enez-Reyes S.,  Mathur  S. et al.: 2008b, {\it Astron. Nachr.}, in press 
\bibitem{} Gelly B., Lazrek M., Grec G. et al.: 2002, {\it A\&A} {\bf 394},  285
\bibitem{} Isaak, G.: 1982, {\it Nature} {\bf 296}, 130
\bibitem{} Jim\'enez-Reyes S. J., Chaplin W. J., Elsworth Y. et al.: 2007, {\it ApJ}  {\bf 654}, 1135
\bibitem{} Kumar P., Quataert E. J. and Bahcall J.: 1996, {\it ApJ}  {\bf 458},  L83
\bibitem{}Mathur S.: 2007, Th\`ese de doctorat, Paris XI
\bibitem{}Mathur S., Eff-Darwich, A., Garc\'ia R. A. and Turck-Chi\`eze S.: 2008, {\it A\&A}  accepted
\bibitem{}Mathur S., Turck-Chi\`eze S., Couvidat S. and Garcia R. A.: 2007, {\it ApJ}  {\bf 668},  594
\bibitem{}Nghiem P. A. P.,  Garc\'ia R. A. and  Jim\'enez-Reyes S.: 2006, {\it ESA-SP 624}  70
\bibitem{Tur2006} Turck-Chi\`eze S.: ~2006, {\it Adv. Space Res.}  {\bf 37},  1569
 \bibitem{Tu01a} Turck-Chi\`eze S., Couvidat S., Kosovichev A. et al.: 2001,
{\it ApJ}  {\bf 555},  L69
\bibitem{Tur2004b} Turck-Chi\`eze S.,  Garc\'ia R. A., Couvidat S. et al.:~2004a, {\it ApJ}  {\bf 604},  455
\bibitem{}Turck-Chi\`eze S., Couvidat S., Piau L. and Ferguson, E.:  2004b, {\it Phys. Rev Lett.}  {\bf 211102}  
\bibitem{Tur2005} Turck-Chi\`eze S et al. :~2005, {\it ESA-SP 588}  {\bf 39th ESLAB symposium}  p 193
 \bibitem{Tur2006} Turck-Chi\`eze S., Carton, P. H., Ballot, J. et al.:~2006, {\it Adv. Space Res.}  {\bf 38},  1812 
 \bibitem{Tur2008} Turck-Chi\`eze S. \& Talon, S.: 2008, {\it Adv. Space Res.}   {\bf 41},  855 
 \bibitem{Tur2008} Turck-Chi\`eze S et al.:~2008, {\it Experimental Astronomy}  The DynaMICCS mission, submitted 


\end{thebibliography}
\end{document}